\documentclass{kapproc} % Computer Modern font calls

\normallatexbib

\def\p{\partial}
\def\re{\mbox{e}}
\def\be{\begin{equation}}
\def\ee{\end{equation}}
\def\bdm{\begin{displaymath}}
\def\edm{\end{displaymath}}
\def\bea{\begin{eqnarray}}
\def\eea{\end{eqnarray}}
\newcommand{\rd}{\mbox{d}}
\newcommand{\ri}{\mbox{i}}

\begin{document}

\articletitle{Wave Functions Statistics at Quantum Hall Critical Point}

%\articlesubtitle{This is an Article Subtitle}

\author{A. M. Tsvelik}
\affil{Department of  Physics, Brookhaven\\
National Laboratory, Upton, NY 11973-5000, USA}
\email{tsvelik@bnl.gov}

%\author{Second Author}
\affil{Contribution to the proceedings of the\\
NATO Advanced Research Workshop on Statistical Field Theories, Como 18-23 June, 2001}
%\email{secondauthor@anotheruniv.edu}

\begin{abstract}
 I elaborate on the earlier suggestion  that the model describing the plateaux transition in Integer Quantum Hall effect scales to a particular point on the line of critical points of a theory with a higher symmetry.
\end{abstract}


\section*{Introduction}

 Under the conditions of low temperature and strong perpendicular
magnetic field, a two-dimensional electron gas exhibits a striking
macroscopic manifestation of a quantum phenomenon, namely the quantum Hall effect
\cite{VonKliz:iqhe,Tsui:iqhe}: the Hall conductivity exhibits quantized
plateaus at well defined multiples of $e^2/h$ (a fundamental constant). In not too clean samples, where the r{\^o}le of
random impurities (disorder) is more important than
electron-electron interactions, the plateaus occur at integer
multiples of $e^2/h$ giving rise to the so-called integer quantum Hall
effect (IQHE). It is widely believed that in the absence of a magnetic
field, all wavefunctions for non-interacting, disordered electrons in
two dimensions are localized. In the presence of a magnetic field
however,  a delocalized state  occurs at the centre of the (disorder
broadened) Landau level, with energy $E_c$. As one tunes the electron
energy $E$ (by varying the magnetic field), through the centre of a
Landau level, the localization length, $\xi$, diverges
as $\xi=|E-E_c|^{-\nu}$, where numerical simulations indicate that $\nu\sim
2.3$.\footnote{We note, that in a system of size $L$, the number of
states with wave functions that reach the boundaries is $\sim
\rho(E)L^{2-1/\nu}$. Since the density of states, $\rho(E)$, remains a
smooth function of energy, this number is always macroscopic. However,
the density of delocalized states $\sim L^{-1/\nu}$ and goes to zero in
the limit of infinite sample size.} The plateaus with differing $\sigma_{xy}$ are  separated  by these critical points.  A theoretical
description of these points remains one of the
most challenging unresolved problems in the theory of disordered systems.

 The supersymmetric sigma model describing disordered Landau levels has a high symmetry. So we are lead to believe that the model describing the critical point has at least that symmetry or even higher. This constitutes a difficulty because on one hand all  known critical theories  with internal symmetry have at least one free parameter (for instance, for Wess-Zumino-Novikov-Witten (WZNW) models it would be  the level) and , on the other hand, there is no such free parameter for  IQHE critical point.    
 In our previous paper we concentrated on properties of LDOS since  this is a much 
 simplier problem than a  calculation of conductances. With LDOS one can work with a closed system and not to   worry about boundary conditions. 

 The first assumption of our work is that the operators representing 
 LDOS constitutes a decoupled sector of the theory. Therefore though the system in its entirety is described by some  (yet unknown) supersymmetric theory, the  
 LDOS sector is much simplier.  This assumption was based on the research conducted for one-dimensional limit of the relevant supersymmetric sigma  model, 
 where the decoupling of LDOS sector was demonstrated explicitely  \cite{Efetov:super}, \cite{mirlin}. These  results  led us to conclusion that the most likely candidate for the theory describing critical LDOS was the H$^+_3$ WZNW model.

\section{H$^+_3$ theory and multifractality}

 $H_3^+$ WZNW  model  is defined on a non-compact space SL(2,C)/SU(2). Since this space differs from the group space SL(2,R) only by the signature of its metric tensor (it is (+,+,+) in the former case and (+,+,-) in the latter one), it is legitimate to consider it as a WZNW model and not a coset one.

In the Gaussian parametrization the action can be written as 

\begin{equation}
\label{slsuaction}
S=\frac{(k + 2)}{4\pi}\int d^2x \left[4\partial \theta{\bar\partial}\theta+e^{2\theta}\partial\mu{\bar\partial}\mu^{\ast}\right].
\end{equation}

 In the  semiclassical limit  $k >> 1$ the primary fields satisfy the diffusion equation on  the H$^+_3$ space:
\begin{equation}
\label{sleigen}
\frac{1}{k}\left(
Y^2\frac{\partial^2}{\partial Y^2}+
Y\frac{\partial^2}{\partial\mu^\ast\partial\mu}\right)\Phi(\tau; \mu,
\mu^\ast, Y)= - \frac{\p}{\p\tau} \Phi(\tau;\mu, \mu^\ast, Y). 
\end{equation}
where $\tau = \ln z$ and $Y = \exp(- 2\theta)$.

 This equation provides a  natural link between the 2D WZNW model and the work done on one-dimensional sigma model. It was found that the eigenfunctions of the corresponding transfer matrix satisfy a similar equation:
\begin{equation}
\label{sleigen}
\frac{1}{k}\left(
Y^2\frac{\partial^2}{\partial Y^2} - 
Y\right)W(\tau; Y)= \frac{\p}{\p\tau}W(\tau; Y). 
\end{equation}
where  the $q$-th power of LDOS $\rho^q$ was  identified as $Y^{q}$. This identification naturally follows  from the form of the correlation function of LDOS derived in \cite{mirlin}. 

 The relationship between $W$ and $\Phi$ is obvious:
\bea
W(Y) \sim \int \rd^2\mu \exp[\ri (k\mu^* + k^*\mu)]\Phi( \mu,
\mu^\ast, Y), ~~ kk^* = 1\label{W}
\eea
To identify  the operator $\rho^q$ one has to recall some facts from the representation theory. 

The primary fields $\Phi^j(x|z)$ of the H$^+_3$ model belong to representations of the SL(2,R) group with angular momentum $j$. The auxilary coordinate $x$ parametrizes the action of generators of the algebra sl(2) represented as differential operators:
\bea
{\cal D}^+_j = -x^2\p_x + 2jx, ~~{\cal D}^0_j = - x\p_x + j, ~~ {\cal D}^-_j = - \p_x
\eea
The coordinates $x,\bar x$ parametrize a direction of the quantization axis; this is reflected in the field parametrization used in the majority of papers on the  H$^+_3$ model:
\bea
\Phi^{(j)}( \mu,
\mu^\ast, Y = \re^{-2\theta}) = \left(|\mu - x|^2\re^{\theta} + \re^{-\theta}\right)^{2j} \label{standard}
\eea
Since these eigenfunctions depend on $(\mu - x)$, one can replace in Eq.(\ref{W}) the integration over $\mu$ by the integration over $x$ and identify:
\bea
W_j(Y) = \Gamma(-2j)\int \rd^2 x \exp[\ri (k\bar x + k^* x)]\Phi^j(x|z), ~~ kk^* = 1
\label{W1} 
\eea 
where semiclassically 
\bea
W_{j}(Y=\re^{-2\theta}) = 2\re^{-\theta}K_{(1 +2j)}(2\re^{-\theta}) \label{W2}
\eea

 To establish a relationship between field $\exp(2q\theta)$  and (\ref{standard}), let us  recall that the scaling  dimensions of H$^+_3$ primaries are 
\bea
d(j) = - \frac{2j(j + 1)}{k}
\eea
where $k + 2$ is the level of the corresponding Kac-Moody algebra. 
As is known from the numerical simulations, the scaling dimension the $q$-th power of LDOS fits to this formula with  $j = -q$ and $2/k \approx 0.262 \pm 0.003$ \cite{mirlin2}. This publication   has also explained that the 
 deviations from the parabolic dependence reported in the earlier work \cite{Janssen:mulfrac} had their origin in  the  finite size effects. 

 The fact that the corresponding  value of $k \approx 7.63$ is not an integer  does not constitute a problem for the $H_3^+$ theory,  since in non-compact models $k$ is not quantized. However, this probably means that the parent supersymmetric  theory which includes in itself a compact sector, cannot possess Kac-Moody symmetry. This supports the point of view of  Zirnbauer who rejected Kac-Moody symmetry for IQHE critical point  \cite{martin}. 

 So we suggest  that $\rho^q$ is related to the primary field with $j = -q$. To extract it from $\Phi^{(-q)}(x)$ is the same as to extract  $\exp(2q\theta)$ from expression (\ref{standard}), which can be done by integration around a contour encircling infinity:
\bea
\exp(2q\theta) = \int_{C}\rd x\int_{C}\rd \bar x x^{(2q -1)}{\bar x}^{(2q -1)}\Phi^{(j = -q)}(x,\bar x,z, \bar z)
\eea

\section{Zero dimension operator}
 In our previous paper we argued  that in order to reproduce existing results on correlation functions of LDOS, one has to consider a certain modification of the  $H^+_3$ WZNW model. Namely,  $n$-point  correlation functions of LDOS should be understood as $n+2$-point function with two additional operators $\psi_0$ present at zero and infinity. In cylindrical geometry these points become minus and plus infinity respectively. The  operator  $\psi_0$ was identified as a non-trivial operator with zero conformal dimension.

 Primary fields of $H^+_3$ WZNW model have been classified and their fusion rules are known. Therefore we have to find $\psi_0$ among these operators. It turns out that  such non-trivial (that is non unity) operator  can be identified as  
\bea
\psi_0(x|z) = \lim_{j \rightarrow 0}j^{-1}\Phi^j(x|z)
\eea
This limit is well defined  semiclassically as follows from  relation (\ref{W2}).  To make sure it holds in quantum field theory, one can study the fusion rules.  Let us consider a three-point correlation function of primary fields in the $H^+_3$ model \cite{teschner1},\cite{teschner2}:
\bea
&&<\Phi^{j_1}(x_1|z_1)\Phi^{j_2}(x_2|z_2)\Phi^{j_3}(x_3|z_3)> = \nonumber\\
&&\frac{C_{123}D_{123}}{|z_{12}|^{d_1 + d_2 - 2d_3}|z_{13}|^{d_1 + d_3 - 2d_2}
|z_{23}|^{d_2 + d_3 - 2d_1}} \label{three}
\eea
where the structure constant $C_{123}$ is  the Clebsh-Gordon coefficient  for the SL(2,R) group.
\bea
C_{123} = |x_{12}|^{2(j_1 + j_2 - j_3)}|x_{13}|^{2(j_1 + j_3 - j_2)}|x_{23}|^{2(j_2 + j_3 - j_1)}
\eea
and $D$ is the quantum correction given by 
\bea 
&&D_{123} = \label{quant}\\
&&\frac{\lambda^{j_1 + j_2 + j_3 +1}{\cal Y}_W(2j_1 +1){\cal Y}_W(2j_2 +1){\cal Y}_W(2j_3 +1)}{{\cal Y}_W(j_1 + j_2 + j_3 +1){\cal Y}_W(j_1 + j_2 - j_3){\cal Y}_W(j_1 + j_3 - j_2){\cal Y}_W(j_2 + j_3 - j_1)} \nonumber
\eea
where 
\[
\lambda = \pi k^{-1/k}\frac{\Gamma(1 - 1/k)}{\Gamma(1 + 1/k)}
\]
The function  ${\cal Y}_W(j)$ is a meromorphic function introduced in \cite{zz}. It has  zeroes at $j = n + mk, n,m = 0,1,2,...$ and $j = - (n +1) - (m+1)k, n,m = 0,1,2,...$ At $0 >  \Re e j > - (1 + k)$ this function admits the integral representation
\bea
&&{\cal Y}_W(- x) = \exp\left\{\int_0^{\infty}\frac{\rd t}{t}\left[k^{-1}(\frac{1+k}{2} - x)^2\re^{-t} - \frac{\sinh^2(1 + k - 2x)t}{\sinh 2t\sinh 2kt}\right]\right\}, \nonumber\\
&&0 <   \Re e x <  (1 + k) 
\eea
Outside this interval it can be defined using the following properties:
\bea
{\cal Y}_W(j) = {\cal Y}_W(j -1) \frac{\Gamma(1 + j/k)}{\Gamma(-j/k)}, ~~{\cal Y}_W(j) = {\cal Y}_W(j - k) k^{-(2j + 1)}\frac{\Gamma(1 + j)}{\Gamma(-j)}
\eea
 Since at $k \rightarrow \infty$ the function $D_{123} = 1$, formula (\ref{three}) has a simple semiclassical limit corresponding to the quantum mechanics of a free particle on the $H_3^+$ space.
 The  property crucial for our argument is the fact that structure constant $D$ contains the product of  ${\cal Y}_W(2j + 1)$-functions  which vanishes when one of the angular momenta goes to zero. Since ${\cal Y}_W(2j + 1) \sim j$ at $j \rightarrow 0$, we get from (\ref{three}):
\bea
&&\int \rd^2 x_2 |x_2|^{(4q - 2)} <\Phi^{j_1}(x_1|z_1)\Phi^{-q}(x_2|z_2)\Psi_0(0|0)> = \nonumber\\ 
&&|x_1|^{2(j_1 + q)}\frac{D(j_1,-q)}{|z_{12}|^{d_1 + d_2}|z_1|^{(d_1 -2d_2)}|z_2|^{(d_2 -2d_1)}} \label{three1}
\eea
\bea
D(j_1,-q) = \frac{\lambda^{j_1 -q  +1}{\cal Y}_W(2j_1 +1){\cal Y}_W(-2q +1)}{{\cal Y}_W(j_1 -q  +1){\cal Y}_W(j_1 -q){\cal Y}_W(j_1  + q){\cal Y}_W(-q  - j_1)} \label{quant1}
\eea
When two operators have $j \rightarrow 0$ we get:
\bea 
&&<\Psi_0(x_1|0)[\rho]^q(z)\Psi_0(x_3|\infty)> = |x_{13}|^{2q}A(q)\nonumber\\
&&A(q) = \frac{\lambda^{-q  +1}{\cal Y}_W(-2q +1)}{{\cal Y}_W(-q  +1){\cal Y}_W^2(-q){\cal Y}_W( q)}
\eea
As it was mentioned above, we suggested that the  two-point disorder average  of $q$-th powers of LDOS should be understood as  the four-point function of the $H_3^+$ theory. In the cylindrical geometry where $z = \exp(w/R)$ I get the following expression:
\bea
&&\overline{[\rho]^q(w_1)[\rho]^q(w_2)} = \int \rd^2 x \re^{\ri{\bf k x}}\times\nonumber\\
&&\int<\Psi_0(\Re e w \rightarrow -\infty|0)[\rho]^q(w_1,\bar w_1)[\rho]^q(w_2,\bar w_2)\Psi_0(\Re e w \rightarrow +\infty|x)> = \nonumber\\
&&|2R\sinh(w/2R)|^{- 2d_q}\int \rd^2 x \frac{|x|^{4q}}{|x - 1|^2}{\cal F}(x,\bar x;z,\bar z), ~~z = \exp(w_{12}/R)
\eea
To derive the latter formula I used the fact that the confomal blocks depend 
only on the anharmonic ratios of $x_i$ and $z_i$. Function ${\cal F}$ satisfies the following differential equation:
\bea
&&\frac{x(1 - x)(1 - xz)}{z(z - 1)}\p_x^2{\cal F} + \nonumber\\
&&\left[\frac{(1 - x)^2}{(z - 1)} + \frac{2x(1 -q) -1}{z}\right]\p_x{\cal F} + k\p_z{\cal F} = 0
\eea
 At the moment I am still unable to present a complete solution of this equation.

 I  acknowledge the support from 
 US DOE under contract number DE-AC02 -98 CH 10886.

\begin{acknowledgments}
I  acknowledge the support from 
 US DOE under contract number DE-AC02 -98 CH 10886. I am grateful to A. Mirlin And F. Evers for the important exchange of opinions. 

\end{acknowledgments}

\end{document}